\documentclass[aps,pre,preprint,groupedaddress,showpacs]{revtex4}
\usepackage{latexsym}
\usepackage{graphicx}
\begin{document}

\title{A Quantum Mechanical Approach To The Polarization Transport of Photons}

\author{Reza Torabi}
\affiliation{Physics Department, Tafresh University, Tafresh, Iran}
\email{rezatorabi@aut.ac.ir}

\begin{abstract}
Based on quantum mechanical approach the polarization transport of
photons which propagate in a medium with slow varying refractive
index is studied. The photon polarizations are separated in opposite
directions normal to the ray which is called "Spin Hall effect" of
photons, and also the rotation of polarization plane, a
manifestation of the Berry phase, occurs. This approach can be
generalized to other spinning particles in inhomogeneous media as a
universal approach.
\end{abstract}


\maketitle

\section{Introduction}

In recent years progressively increasing attention is given to the
phenomena of topological spin transport of quantum particles (see,
for instance
\cite{Torabi,Mehrafarin,Murakami,Sundaram,Culcer,Bliokh,Bliokh2,Bliokh3,Bliokh4}
and references therein). These effects are closely related to the
general notion of Berry phase \cite{Berry} and thus should also be
showed up in the electromagnetic wave transport (so-called "Spin
Hall effect of photons"). It can lead to the topological splitting
of a ray of mixed (non-circular) polarizations into two circularly
polarized rays due to the inhomogeneity of medium \cite{Bliokh2}. As
it was demonstrated in \cite{Bliokh3,Onoda}, this effect is
completely analogous to the spin Hall effect in solids (see
\cite{Murakami,Culcer}).

The spin Hall effect of photons has been described based on the wave
packet dynamics \cite{Onoda} and post geometric optics approximation
\cite{Bliokh2} both in inhomogeneous media with slow varying
refractive index. In this paper we introduce a quantum mechanical
approach to these media. The priority of this approach is that it
can be generalized to other spinning particles in inhomogeneous
media as a universal approach.

The different polarizations, left and right hand polarizations, are
degenerate in a homogeneous isotropic medium
\cite{Bliokh3,Kravtsov}. In our case this double degeneracy is
lifted by an additional term in Hamiltonian which is interpreted as
spin-orbit interaction of photons and leads to the Berry phase and
Berry curvature. The Berry curvature in momentum space of photons
with opposite helicities has the form of opposite-signed 'magnetic
monopoles' located at the origin of the momentum space. The Berry
phase gives rise to the polarization evolution law \cite{Rytov} and
the Berry curvature causes an additional effective force which
deflects the rays in opposite directions depending on their
polarization. The latter phenomenon is a manifestation of the
topological spin transport or the Spin Hall effect of photons.

This paper is organized as follow. In section II, the photon
Hamiltonian in inhomogeneous media is presented. In order to obtain
equations of motion, the Hamiltonian is diagonalized  in section III
by using a unitary transformation. During this procedure a gauge
potential appears which is diagonal in the helicity basis in
adiabatic approximation. Section IV is allotted to the derivation of
semiclassical equations of motion.

\section{Hamiltonian in Inhomogeneous Media}
An isotropic medium turns out to be weakly anisotropic due to the
presence of a selected direction determined by a gradient of the
refractive index. The change in refractive index affects the
direction of wave vector ${\bf k}$ or equivalently the direction
of photon's momentum ${\bf p}$. Slow variation of refractive index
guarantees the adiabatic condition \cite{Messiah} which renders
helicity, polarization, as an adiabatic invariant \cite{Chiao}.

In the absence of polarization we can construct a wavepacket
$|\psi\rangle$, which represents a point particle (photon), by the
states $|p\rangle$ in the momentum space. When we consider the
polarization degrees of freedom, the states are
$|p,\pm\rangle=|p\rangle\otimes|\pm\rangle$ and the constructed
wavepacket represent a particle with a spin (polarization). The
Hamiltonian in this case will be
\[
H({\bf p},{\bf x})=H_0({\bf p},{\bf x})+H_s,
\]
where $H_0$ is the Hamiltonian of the point particle in weakly
anisotropic media and $H_s$ is the contribution of spin
(polarization) term. As was mentioned, the direction of photon's
momentum varies during the photon's trajectory, so the Hamiltonian
$H_0$ depends on ${\bf p}$ and ${\bf x}$(explicitly through ${\bf
p}={\bf p}({\bf x})$).

The spin of the photon always follows the direction of ${\bf p}$ and
the evolution of it is governed by the Hamiltonian $H_s$ as
\cite{Chiao}
\[
H_s=\kappa{\bf \sigma}\cdot {\bf p},
\]
where the constant $\kappa$ is to be determined by experiment. The
general Hamiltonian will be
\begin{equation}
\label{eq1} H({\bf p},{\bf x})=H_0({\bf p},{\bf x})I
+\kappa\sigma\cdot {\bf p},
\end{equation}
where $I$ is a $2\times2$ unit matrix. Double degeneracy of the
polarization, left hand and right hand, is lifted by the additional
term, second term in the above equation, which is spin-orbit
interaction of photons.

\section{Diagonalization of Hamiltonian in Adiabatic Approximation}

In order to derive equations of motion, we first diagonalize the
Hamiltonian (1). The non-diagonal part of the Hamiltonian is the
$H_s$.

The photon's momentum is ${\bf p}(t)=p(t){\bf n}$ where ${\bf
n}=(sin\theta cos\varphi,sin\theta sin\varphi,cos\theta)$ is the
unit vector and $\theta$ and $\varphi$ are zenithal and azimuthal
angles, thus the Hamiltonian can be diagonalized with the unitary
transformation
\begin{equation}
\label{eq2}
U(p)=exp(-\frac{i\theta\sigma_2}{2})exp(-\frac{i\varphi\sigma_3}{2}),
\end{equation}
in which $U^\dag(p)(\sigma\cdot p)U(p)=p\sigma_3$. Eigenmodes of
$\sigma_3$ physically describe the helicity quantum number
$\lambda=\hbar^{-1}\frac{{\bf \sigma}\cdot {\bf p}}{p}$. Under this
unitary transformation the new Hamiltonian
$\tilde{H}=U^\dag(p)HU(p)$ becomes
\[
\tilde{H}=H_0({\bf p},{\bf x}I+{\bf A})+\kappa p\sigma_3,
\]
where
\[
{\bf A}({\bf p})=i U^\dag(p)\nabla_{\bf p} U(p).
\]
In deriving the above, we have made use of the identity (see
 Appendix A)
\begin{equation}
\label{eq3}
G^\dag(x)f(\partial_x)G(x)=f(\partial_x +G^\dag
\partial_x G).
\end{equation}
Direct calculation via equation (\ref{eq2}) yields the following
expression for the components of the vector matrix ${\bf A}$
\[
A_k=0,{\;\;\;} A_\theta=\frac{1}{p}\sigma_2,{\;\;\;}
A_\varphi=\frac{1}{p}[cotg\theta \sigma_3-\sigma_1].
\]
The unitary transformation induces the non-Abelian gauge potential
which is a pure gauge one,i.e., the corresponding field strength,
$F_{ij}=\partial_{p_i} A_j - \partial_{p_j} A_i + i[A_i,A_j]$, is
identically zero. The canonical coordinate conjugate to ${\bf p}$ is
${\bf x}$ which corresponds to the usual derivative
$i\hbar\nabla_{\bf p}$. In the absence of ${\bf A}$, this canonical
coordinate is the physical (observable) coordinate. In the presence
of ${\bf A}$, however, the physical coordinate becomes
\[
{\bf r}=U^\dag(i\hbar\nabla_{\bf p})U={\bf x}I+\hbar{\bf A}.
\]
So that it now corresponds to $i\hbar D_{\bf p}$, where $D_{\bf
p}$ is the covariant derivative defined by
\[
D_{\bf p}=I\nabla_{\bf p}-i{\bf A}.
\]

As was mentioned, helicity is invariant in the adiabatic
approximation in which the diagonal components of ${\bf A}$ are
retained. Thus ${\bf A}$ will be a $2\times2$ diagonal matrix in the
helicity basis as follow
\[
{\bf A}=p^{-1} cotg(0,0,1)\sigma_3.
\]
In this representation, the gauge potential (Berry connection) is
$\lambda{\bf A}$, where $\lambda=\pm1$ and {\bf A} is now a vector.
The field strength (Berry curvature) becomes $-\lambda\frac{{\bf
p}}{p^3}$ which is the field of a magnetic monopole of charge
$-\lambda$ situated at the origin of the momentum space.
Furthermore, the physical coordinate reduces to
\begin{equation}
\label{eq4} {\bf r}={\bf x}+\lambda\hbar{\bf A}.
\end{equation}

\section{Berry effect in the semiclassical equations of motion}

Let's consider how the presence of the gauge potential ${\bf A} $
affects the commutation relations of physical variables and the
equations of motion. We have from (\ref{eq4}) that
\[
[r_i ,r_j]=i \lambda\hbar^{2} \varepsilon_{ijk} \frac{p_k}{p^{3}},
\]
Other commutation relations are $[p_i ,p_j]=0$ and ${\;\;}[r_i
,p_j]=i\delta _{ij}$. In the other words the space will be
noncommutative \cite{Berard}. Similar commutation relations are
achieved in \cite{Pati} by parallel transport of photon wave vector
in momentum space and also an Aharonov-Bohm effect was predicted. In
this paper ,by using the quantum mechanical approach, we show that
the rotation of polarization plane which is a manifestation of Berry
phase, and also the spin Hall effect of photons occur in
inhomogeneous media. The Heisenberg equations have the usual
canonical form in the generalized variables
\[
\dot{\bf p}=-\frac{i}{\hbar }[{\bf p}, \tilde{H}]\quad , \quad
\dot{\bf r}=-\frac{i}{\hbar}[{\bf r},\tilde{H}],
\]
where the photon Hamiltonian is taken as a function of generalized
variables, $\tilde{H}({\bf p},{\bf r})$. Using the commutation
relations, the semiclassical equations of motion will be
\begin{equation}
\label{eq5} \dot{\bf p}=-\nabla_{\bf r} \tilde{H} \quad , \quad
\dot{\bf r}=\nabla_{\bf p} \tilde{H} +\lambda\hbar(\frac{{\bf
p}}{p^3}\times {\dot{{\bf p}}}),
\end{equation}
up to the first order in $\hbar$. These relations reduce to the
standard ray equations of geometric optics in the classical limit
$\hbar\rightarrow 0$ or in the absence of polarization $\lambda=
0$. The second term in the right-hand side of equations
(\ref{eq5}) constitutes the corrections that describe topological
spin transport of the photon. We notice that this term causes an
additional displacement of photon of distinct helicity in opposite
directions normal to the ray.

As a consequence, the magnitude of splitting for the rays of
left-hand and right-hand polarizations is determined by a contour
integral in the momentum space
\[
\delta {\bf r}=\lambda \int_C \frac{({\bf p}\times d{\bf p})}{p^3},
\]
where $C$ is the contour in the $p$-space along which the photon
moves. Hence the Spin Hall effect of photons is a nonlocal
topological effect.

The resulting geometric Berry Phase, $\sigma\int_C{{\bf A}\cdot
d{\bf p}} $, has opposite signs for the two polarizations.
Therefore, for a linearly polarized wave, this Berry phase leads to
the rotation of the polarization plane through the angle
\begin{equation}
\label{eq6} \gamma=\int_C{{\bf A}\cdot d{\bf p}}=\int_C{cos \theta
d\varphi}.
\end{equation}
This is the Rytov law for photons, a manifestation of the Berry
phase, which is topological in nature as the spin Hall effect is.

The above finding for Berry effect, equations (\ref{eq5}) and
(\ref{eq6}), coincide with the result \cite{Bliokh2,Onoda}, which
were obtained differently.

\section{summery and conclusion}

By using a quantum mechanical approach, it was shown that if we
consider the spin degrees of freedom of photon in media with slow
varying refractive index, an abelian gauge field will be appeared in
photon position operator. This gauge field causes the space to be
non-commutative which leads to an additional displacement of photon
of distinct helicity in opposite direction normal to the ray (spin
Hall effect of photons). Also the rotation of polarization plane
occurs which is topological in nature as the spin Hall effect is.
The presented quantum mechanical approach can be applied to other
spinning particles in inhomogeneous media as a universal approach.
\[
\]
\textbf{Appendix A: Appearance of pure gauge potential}

To prove the identity (\ref{eq3}), we expand the operator
$f(\frac{d}{dx})$ in taylor series, therefore we deal with
operator of the form $(\frac{d}{dx})^n$, $n\epsilon N$. If we
demonstrate that
\begin{equation}
\label{eqA1} U^\dag(x)(\frac{d}{dx})^n U(x)=(\frac{d}{dx}+U^\dag
\frac{dU}{dx})^n,
\end{equation}
then the identity will be proved. for this purpose we use priori.

For $n=1$ the expression is true: $U^\dag(x)\frac{d}{dx}
U(x)=\frac{d}{dx}+U^\dag \frac{dU}{dx}$. If equation (\ref{eqA1}) is
true for some $n=i$ then, for $n=i+1$ one has
\[
U^\dag(\frac{d}{dx})^{i+1} U=U^\dag(\frac{d}{dx})^i \frac{d}{dx}
U=U^\dag(\frac{d}{dx})^i U \frac{d}{dx}+U^\dag(\frac{d}{dx})^i
\frac{dU}{dx}
\]
\[
=U^\dag(\frac{d}{dx})^i U \frac{d}{dx}+U^\dag(\frac{d}{dx})^i
UU^\dag \frac{dU}{dx}=U^\dag(\frac{d}{dx})^i U
(\frac{d}{dx}+U^\dag \frac{dU}{dx})
\]
\[
=(\frac{d}{dx}+U^\dag \frac{dU}{dx})^{i+1}=(\frac{d}{dx}+U^\dag
\frac{dU}{dx})^{i+1}.
\]
Thus, by periori, the equality (\ref{eqA1}) is proven for any power
of $(\frac{d}{dx})$ and hence, for any analytic function of it.
Three dimensional generalization of equation (\ref{eq3}) can be
proved in a similar way.

\textbf{Acknowledgement}

The author is grateful to A.Chakhmachi for fruitful discussions.

\end{document}